# Towards high-resolution a

**Craig Mackay** reports from a recent meeting on "The future of high-resolution imaging in the visible and infrared", reviewing the astronomical drivers for development and the technological advances that might boost performance.

Substantial developments in imaging technologies in recent years are opening up exciting new opportunities for astronomers. These new technologies will complement the capacity of the next generation of space telescopes and help to deliver new science. We can now image a small number of exoplanets directly. We can trace the dynamics of stars around the black hole in the centre of our galaxy. With gravitational lensing, we can study the structure of very distant galaxies. Fast, low-noise, wide-area detectors in the visible region will revolutionize the detection of rocky planets around Sun-like stars with GravityCam, a proposed UK instrument for the European Southern Observatory's (ESO) New Technology Telescope at La Silla; it will deliver a resolution 2.5 to 3 times better than natural seeing and, with low-order adaptive optics (AO), allow much fainter reference stars to be used. These examples are just some of the imaging applications that prompted the RAS Specialist Discussion Meeting, which was held in London on 8 February 2019 and organized by Colin Snodgrass (University of Edinburgh), Craig Mackay (University of Cambridge), Tim Morris (University of Durham) and Jesper Skottfelt (Open University). This article provides a synopsis of the presentations.

**Ric Davies** (Max Planck Institut für extraterrestrische Physik) spoke on "A clear view of the universe: opportunities for high-resolution imaging". Achieving the highest spatial resolutions from the ground inevitably involves AO or even interferometry. Since the first astronomical demonstration, AO has advanced both in terms of what can be achieved and what is expected (Davies & Kasper 2012). We can expect another leap when ESO's 39 m Extremely Large Telescope (ELT) is completed in 2025. Remarkably, higher resolutions are already being achieved with optical/IR interferometry, which can now map the surfaces of stars. But it is typically technically limited to bright

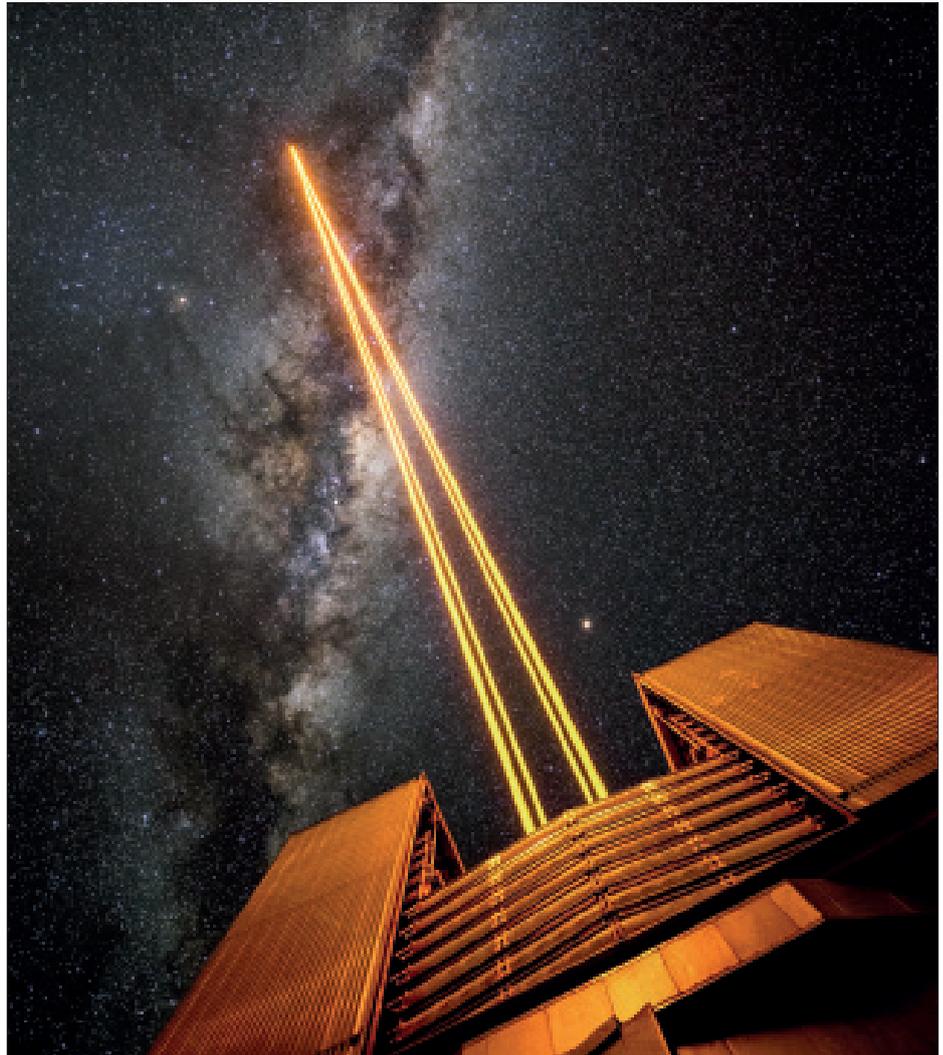

**1** Four lasers create an artificial guide star from Unit Telescope 4 of ESO's VLT at the Paranal Observatory, Chile. This is part of the VLT's adaptive optics system, a technology which corrects the blurring effects of the Earth's atmosphere to produce higher resolution images. (G Hüdepohl [atacamaphoto.com]/ESO) targets, hence its applications lie mostly in stellar and circumstellar astrophysics. A recent new runner in this game is GRAVITY, installed at ESO's Very Large Telescope Interferometer (VLTI) and has been trained on Sgr A* and S2 in the galactic centre and on the broad-line region of the quasar 3C 273 (Abuter et al. (GRAVITY Collaboration) 2018) with stunning success (figure 2).

### VLT and ELT

Returning to the VLT and ELT, Davies gave four examples that show what we can expect in terms of high spatial resolution in the coming decade. Galaxy evolution at high redshift is a key topic. Large samples observed with instruments such as KMOS (the K-band Multi Object Spectrograph on the VLT) at seeing-limited resolution have probed scales of 4–5 kpc in many hundreds of galaxies at $z > 1$. With AO (and, in the future, the James Webb Space Telescope [JWST]), we can probe scales of ~500 pc in some of these, enabling us to see the signatures of physical processes that drive mass assembly and structural transformations: inflow in discs, the growth of bulges, star-forming clumps, feedback and quenching. ELT instruments such as HARMONI and MICADO, as well as ALMA (the Atacama Large Millimetre–submillimetre Array), will enable us to reach sub-100 pc scales out to the highest redshifts, yielding a view as detailed as that we can now achieve when looking at Virgo Cluster galaxies. Technically, the challenge is in terms of sky coverage in the deep fields, and in very long integrations over multiple wavebands, both of which mean using AO in less than optimal conditions.

An alternative way to derive the

# stronomical imaging

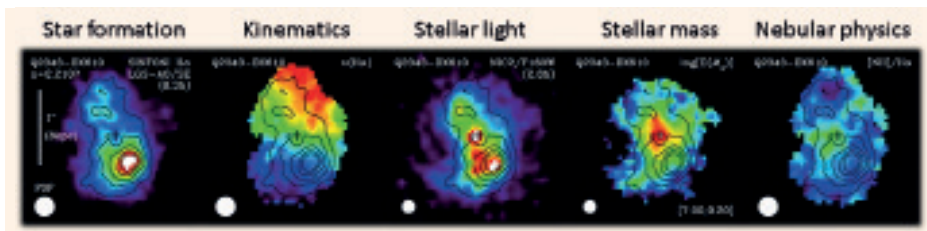

**2** A series of images obtained of quasar Q2343-BX610 at a redshift 2.21.

star-formation history of galaxies at early cosmic times is to spatially resolve the relic populations in local galaxies. The resulting colour–magnitude diagrams provide direct access to the star-formation rates at different cosmic epochs, with the horizontal branch probing ages exceeding 10 Gyr. Crucially, the 6–12 mas resolution of the ELT will enable us to reach to within the central effective radius of galaxies, where the surface brightness is ~20 mag arcsec$^{-2}$. Because stellar populations can be better separated with optical rather than near-infrared colours, there is at the same time a new effort to develop visible AO for the 8m VLT, which serendipitously yields comparable resolution to near-infrared AO on the ELT. The key challenges for this science are achieving the uniform point spread function (PSF) needed for good photometry over the field, and high AO performance at short wavelengths.

Astrometry is a good companion to photometry and the galactic centre is one of the best showcases for such measurements. The orbit of the S2 star around Sgr A* has been followed since 1992, which in itself provides a history of high-resolution imaging. GRAVITY has enabled measurements of the position of S2 with respect to the $K$ ~ 16.6 mag Sgr A* source at a precision of tens of microarcseconds, so that the star is seen to move day by day (Abuter *et al.* 2018). In general, astrometry in dense stellar systems has multiple goals, from measuring anisotropy to derive more robust total masses, to obtaining unambiguous evidence for intermediate-mass black holes. Astrometry is the key challenge here: controlling and calibrating distortions to sufficient precision and characterizing the spatially variant PSF well enough.

High-contrast imaging, in order to detect and characterize exoplanets, was Davies's fourth example. The sweet spot for ground-based telescopes lies in achieving high contrast around bright stars at small radii. For the JWST, with strength in low background noise data, it will lie in the detection of planets around faint primaries at larger radii. Similarly on the ELT, it is the small inner working angle that will be exploited most. Simulations indicate that in just 30 s it will be possible to see the innermost two planets of the HR 8799 system, and that relatively cool planets at even smaller radii would also become detectable. For this work, the technical challenge is to achieve the highest contrasts. This depends on optimizing every part of the optical train and data analysis.

These four science cases that rely on high-resolution imaging show that AO is very much in demand, but also that as we move from the ~50 mas scales now achievable to the ~10 mas scales in prospect with the ELT, there are many challenges to face.

### Adaptive optics
**Norbert Hubin** (ESO) addressed some of those challenges, speaking about the AO systems currently in use on the VLT and in design for the forthcoming ELT – some of the most demanding sets of instruments for large optical telescopes. In the case of the VLT there are major systems for the four large unit telescopes plus others for the VLTI; all of them have very complex technical demands. They will often include visible and near-infrared (NIR) channels, as well as a requirement for atmospheric dispersion correctors (essential for high-resolution work at any significant zenith distance). Many instruments need laser guide-star facilities in parallel with natural guide-star capacity, for basic position offset. For many applications, such as looking at planets around bright stars, one of the key parameters for an AO system is the capacity to deliver high-contrast images. One VLT instrument, ZIMPOL, can achieve contrasts better than one part in 10$^7$.

In recent years, ESO has developed an instrumentation laboratory called the Adaptive Optics Facility (AOF) in Garching bei München, Germany. The following list describes the range of instruments that are now in use on the ESO telescopes or in advanced states of development at the AOF:
● 12 AO systems are supporting VLTI.
● High contrast at 10$^{-6}$ with extreme AO available also in visible light.
● VLT Unit Telescope 4 (figure 1) as an adaptive telescope operating in ground layer, laser tomography and single-conjugate AO modes (GLAO, LTAO, SCAO) and for imaging and spectroscopy in IR and visible (narrow-field mode): pathfinder of ELT.
● Coming soon: laser guide star SCAO second generation in IR imaging and 3D spectroscopy.
● In the pipe: multiple-conjugate AO (MCAO) based visible imager and 3D spectroscopy over 30″ field of view: complementary to the MCAO system in IR for ELT.

Astronomers working with faint targets would benefit greatly from improved image quality on current and planned ground-based telescopes. At present, most AO systems are targeted at the highest resolution using bright guide stars; **Craig Mackay** (University of Cambridge) described low-order AO using very faint reference stars, demonstrating a radically new approach to measuring low-order wavefront errors. After correcting for these errors, significant improvements can be delivered in image resolution in the visible on telescopes in the 2.5–8.2 m range on good astronomical sites. As a minimum the angular resolution may be improved by a factor of 2.5–3 under almost any conditions and, with further correction and image selection, even sharper images may be obtained routinely. Many of the assumptions about what may be achieved with faint reference stars has been re-examined to achieve this performance. Mackay showed how a new design of wavefront curvature sensor combined with novel wavefront-fitting routines allow this performance to become routine. Simulations over a wide range of conditions match the performance already achieved in runs with earlier versions of the hardware described.

In order to use faint reference sources in astronomy it is important to have an approach to wavefront detection and characterization that is flexible and continues to work stably even under conditions of very low light from the reference object. The key way to handle that is to limit the number of Zernike terms that are corrected. Most of the power in atmospheric turbulence

> "ESO has developed an instrumentation lab called the Adaptive Optics Facility"

is in the lowest orders and so the greatest improvement is achieved by correcting for those low orders. Correcting for too many orders simply adds noise and worsens performance. Combining the right approach to wavefront correction with lucky imaging (described below) allows a trade-off between the very sharpest images achieved with a small percentage selection or a slightly less good image profile but with much higher signal-to-noise (figure 3).

### GravityCam

GravityCam (Mackay *et al.* 2018) is a new instrument designed to image large areas of the sky in the visible with angular resolution much better than normally possible with ground-based instruments. **Jesper Skottfelt** (Open University) described this project, which is shared between the Institute of Astronomy at the University of Cambridge, the Open University STEM Faculty with Colin Snodgrass (now at the University of Edinburgh), the Centre for Electronic Imaging under Andrew Holland, and the University of St Andrews under Martin Dominik. GravityCam works by taking images at high speed, typically 25 Hz. Each image is then shifted and combined to give the output image. This technique, known as lucky imaging, improves the resolution by a factor of 2.5–3 over that normally obtained from the ground. Even higher resolution comes from post-processing.

GravityCam is particularly well suited to several scientific programmes. The detection of planets at or below the mass of the Earth is extremely difficult using normal radial velocity or transit methods. GravityCam will make it possible to track up to 90 million stars in the bulge of our galaxy, night after night, to find gravitational microlensing events (figure 4). Should the lensing star have a planet in orbit around it, the microlensing brightness profile is affected in ways that allow astronomers to determine the planet's mass, diameter and orbital radius. We predict that we will detect several thousand new microlensing events over the six-month period when the bulge of our galaxy is in the sky, and expect to detect very many low-mass planets in the region of 0.3–3 au. The sensitivity allows planets with the mass of the Moon to be detected.

Other programmes use the capacity of GravityCam to build statistics on the fluctuations of every star in the field. Microlensing events are selected by abnormal brightening characteristics. These targets are then followed for many weeks. At the same time, we will survey for Kuiper belt and Oort cloud objects with great efficiency. We will see stars blink off and on again for very short periods of time.

GravityCam would use ESO's 3.6 m

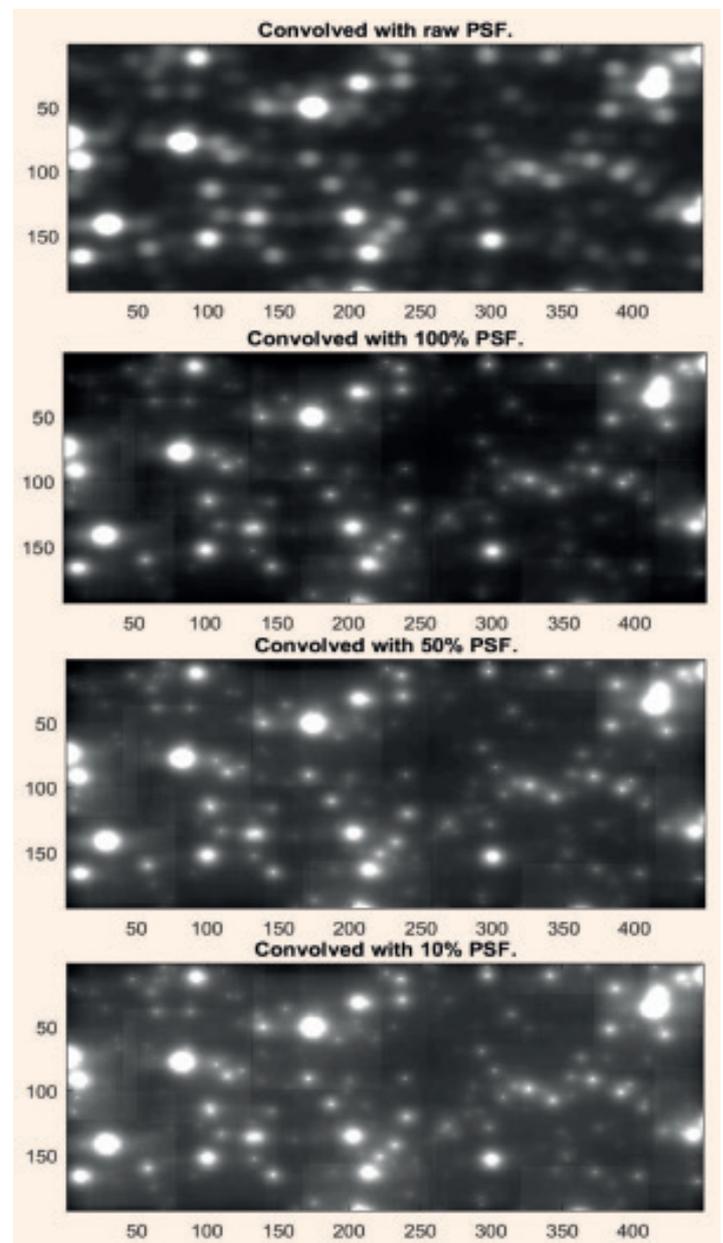

**3** Typical simulation results for a 4.2 m telescope looking at a high-density star field modelled on Baade's window with around 600 stars brighter than $I \sim 25$ in each field of about 10′20as. The pixel scale here is ~45 milliarcsec. It assumes 0.6 as seeing and 10 Zernike terms have been corrected. The core resolution obtained is of the order of 45 mas and is more than twice that of the Hubble Space Telescope. These are characteristic of the appearance of lucky plus low-order adaptive optics images. The central near-diffraction limited core is surrounded by a diffuse halo significantly narrower than the natural seeing.

New Technologies Telescope, a high-performance telescope well-suited to mounting a wide-area detector array in the Naysmith focal plane. The instrument will include an atmospheric dispersion corrector to maintain angular resolution when observing well away from the zenith. The instrument will use wide-area complementary metal–oxide–semiconductor (CMOS) imaging detectors with very high quantum efficiency and low read-out noise. The instrument will produce ~500 TB of data per night, but real-time processing reduces the data rate dramatically.

Building GravityCam does not pose particularly severe technical challenges. The detector manufacturer will probably build the main camera cryostat. Much of the academic effort will be focused on software development to make an efficient and reliable data pipeline. Our estimates are that the total project might cost in the region of $20 million including operations for two years after the end of commissioning, and that it should take about 2.5–3 years to be ready to take data.

**Sasha Hinkley** (University of Exeter) gave an overview of the benefits of including a sparse aperture masking (SAM) capability on the ELT in order to unveil the process of planet formation, and possibly directly detect newly formed massive planets near their formation location. SAM has a host of scientific advantages; it turns the full aperture of any telescope into a collection of interferometers, giving a boost in resolution compared to a conventional telescope. Specifically, the suite of first-generation instruments for the ELT equipped with a SAM customized for observations from 3–5 μm will give access to the circumstellar snow lines (at 2–4 au) where massive, Jupiter-type planets are likely to form. So far, this region of parameter space has been inaccessible, but will become reachable with the superior resolution of a 39 m telescope.

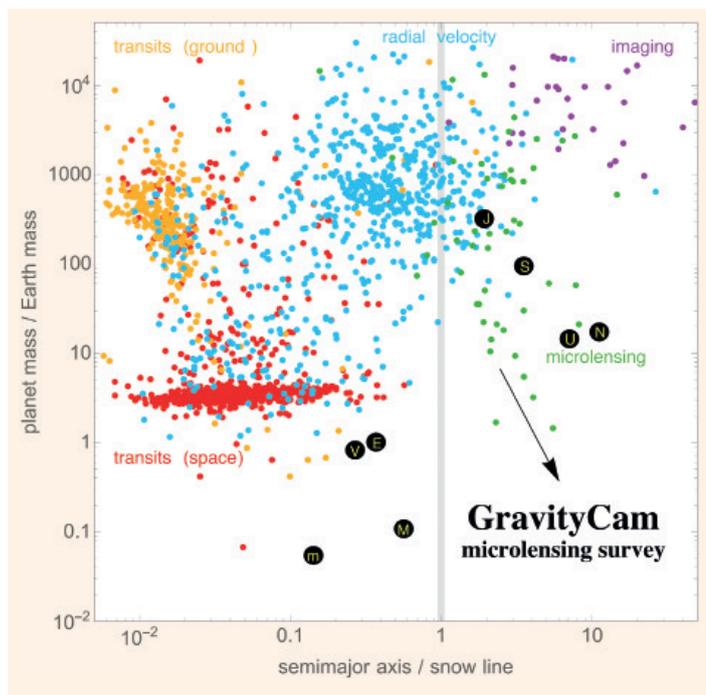

**4** GravityCam will detect planets that are virtually undetectable by other methods. This figure shows the planets detected to date where we see that for planets of Earth-mass and below detections are very few though we expect there are vast numbers of them.

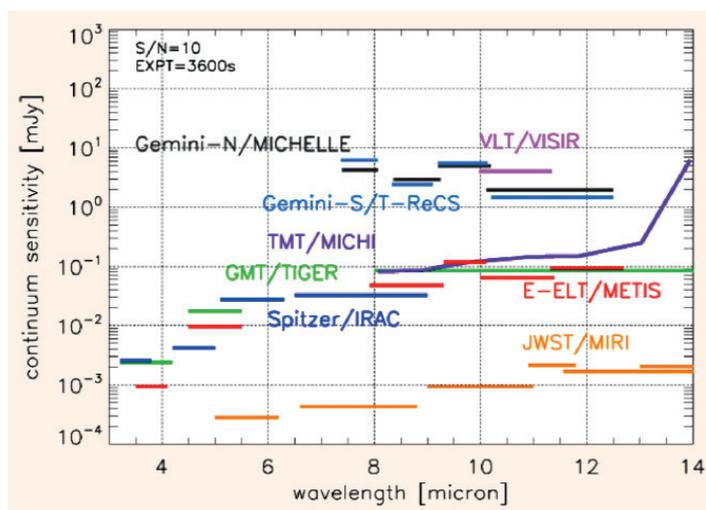

**5** METIS on the ELT has the potential to achieve excellent continuum sensitivity at much longer wavelengths than has been possible so far.

This enhanced resolution in turn provides sensitivity to the snow-line regions for more distant stars, opening the list of potential targets to the nearest star-forming regions, where massive young planets appear brighter and thus easier to detect. Furthermore, the rather different parameter space occupied by SAM will be out-of-reach to the other ELT instruments using conventional imaging modes operating at 1–2 µm. The key instrument is METIS, a first-generation ELT instrument operating in the L/M/N bands (2.9–14 µm) (figure 5). The first two bands may be used with an integral field unit spectrograph to provide resolving power of ~100 000.

High-resolution imaging covers a variety of different applications each of which has significantly different detector requirements. **Kieran O'Brien** (Durham University) spoke about the detectors that will be needed in future and, in particular, the wide range of requirements for AO applications. The science fields of view increase as we move from single-conjugate AO up through ground-layer AO to multiple-conjugate AO. At each stage the demands on the detectors become greater. The science requirements force the use of wider and wider fields of view and, with AO, the angular resolution that can be achieved with a specific telescope is increased, requiring the detector to provide yet more pixels. The wavefront sensors used by these AO systems can also become very demanding; they often require very fast read-out with essentially zero noise.

Other high-resolution techniques such as interferometry and SAM have other requirements. The scientific field of view is generally rather small, but the read-out needs to be fast because the detected images change very rapidly. Interferometry also needs good-quality infrared detectors.

> "One of the most exciting prospects is the microwave kinetic inductance detector"

With lucky imaging, high frame rate is important and, to work at lower light levels, sub-electron read-out noise is essential.

But astronomers and instrument designers are fortunate that the range of detectors currently available is excellent and in many applications specifications are approaching the performance of what we might think of as a perfect detector. For example, 4096 × 4096-pixel high quantum efficiency (>80%) NIR arrays are now in routine use. They depend on a mercury–cadmium–telluride detector layer bump bonded onto a CMOS. This provides fast read-out which for many IR applications is important because the photon rate per pixel can grow quickly.

### Electron-multiplying CCDs

Charge-coupled devices (CCDs) have long been the detectors of choice in visible astronomy. Teledyne e2v (see below) is a major player and over the last decade has introduced devices that have an internal magnification, allowing them to detect individual photons with good signal-to-noise. These electron-multiplying CCDs have been used in lucky imaging as well as fast guiding applications using the CCD 220, which allows kilohertz frame rates over the 220 × 220-pixel frame with significantly less than one electron RMS readout noise.

For future applications one of the most exciting prospects is the microwave kinetic inductance detector (MKID). These are devices that work at very low temperature with superconducting components. They are able to detect individual photon events and have a relatively short recovery time, allowing photon rates to be managed at up to around 1 MHz. MKID structures are now being developed into arrays. The read-out of these devices is unusual and interesting; essentially they involve scanning the detector with a radiofrequency source and detecting the output. Several applications are being examined and the next generation of kinetic inductance detectors is under development in Europe and the US to improve energy resolution, array size uniformity and quantum efficiency. In Europe these developments are going on at Durham, Cambridge, Cardiff, Dublin Institute for Advanced Studies and the Netherlands Institute for Space Research.

**Paul Jorden** (Teledyne e2v) then described developments in visible and IR sensors for astronomical imaging. Teledyne e2v (UK) and Teledyne Imaging Sensors (USA) supply state-of-the-art sensors for astronomical imaging, spectroscopy and related applications. Low noise, high spectral response and generally optimized performance for scientific use are key themes (Jorden *et al.* 2018). He presented examples of standard and custom CCDs up to 9000 × 9000 pixels in multiple formats, some with sub-electron read-noise (enabling photon counting). These are provided as single sensors, sub-assemblies, or complete cryogenic camera systems such as the 1.2 gigapixel J-PAS camera.

A growing component of their portfolio is CIS (CMOS image sensors), which offer performance comparable to the well-established CCDs but with the benefits of considerable digital circuitry for enhanced functionality and immediate image data in digital form. Such sensors are used in ground- and space-based applications including occultation measurements, wavefront sensing and low-power space imagers.

Astronomers regularly require IR sensitivity; the HxRG family of cadmium mercury telluride (CMT) NIR sensors offer formats up to 4000 × 4000 pixels with world-leading performance. While such sensors address the need for NIR sensors at wavelengths beyond 1 µm, the company also develops silicon sensors with increased red sensitivity. CCDs and CMOS imagers operate at wavelengths from X-rays through to 1.1 µm and offer an excellent cross-over with the NIR wavelength range of the CMT sensors (figure 6).

Many of these sensors offer "near-perfect" performance, but there is always room for improvement. New developments are allowing reduced read-noise, better red sensitivity, more varied formats and increasing capability of CIS devices. In particular, it is now feasible to design and construct CMOS mosaics for large field of view and excellent imaging performance.

**David Buscher** (Kavli Institute for Cosmology, Cambridge) then described progress with the Magdalena Ridge Observatory Interferometer (MROI), a next-generation optical/infrared aperture synthesis array under construction in New Mexico (Buscher *et al.* 2013). Its maximum baseline of 350 m allows imaging at angular resolutions down to 0.3 mas, exceeding the resolution of the ELT by more than an order of magnitude. The science case for the MROI rests on the combination of this interferometric resolution with the ability to see faint objects inaccessible to existing interferometers and to image targets much more rapidly (figure 7). Existing interferometers can take several nights of observing to make an image of a single object, but the full-up MROI with its 10 telescopes will be able to make 50 images per night.

The optical instrumentation group in Cambridge (part of Cavendish Astrophysics) is collaborating with New Mexico Tech (NMT) on the design and construction of the array. In the past year there have been several significant developments, chief of which is the installation and commissioning of the first

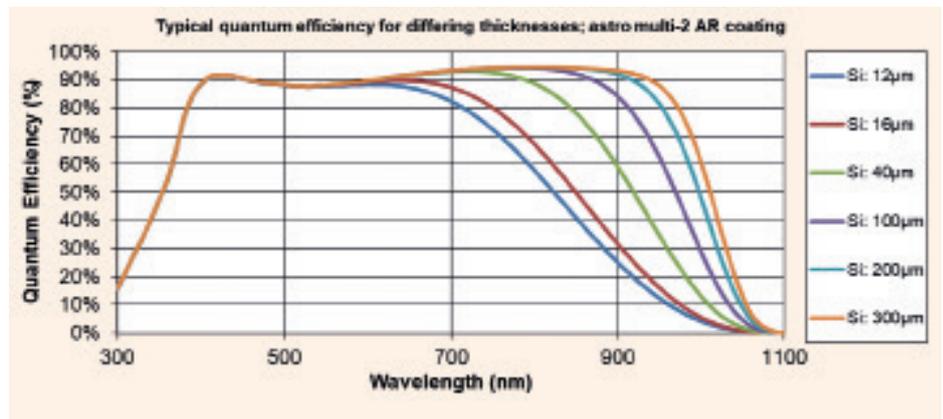

**6** The quantum efficiency of different silicon thicknesses when they are deep depleted. Most conventional sensors (CCD and CMOS) are manufactured with an epitaxial thickness of 12–40 µm. By using thicker non-epitaxial, high-resistivity silicon sensors can be manufactured with greater thickness. This significantly increases the red sensitivity, especially at wavelengths >800 nm. They are back illuminated and antireflection coated to give the highest quantum efficiency over as wide a spectral range as possible.

unit telescope and its enclosure on the array last summer. This was followed by the commissioning of the tip/tilt AO system on the unit telescope by the Cambridge team in November 2018. The commissioning run took three nights, and on all three the seeing was sub-arcsecond, testifying to the intrinsic quality of the site and the absence of significant "dome seeing" contributions from the unit telescope and its enclosure.

These developments have mitigated a significant fraction of the technical risk to the MROI project. In addition, further funding from the US Congress means that the path to the first fringes is now clear. The orders for the second unit telescope and its enclosure have been placed, with delivery targeted for mid-2020. Cambridge is working on delivery of the automated alignment system, the second delay line and the first science instrument, a three-way spectrally dispersed beam combiner, as well as participating in the commissioning of the interferometer. This, together with the ongoing work at NMT on the beam relay system and the fringe tracker, means that the MROI is on course for first fringes in late 2020. It is looking likely that the order for a third telescope could be placed this year, making three-telescope operation possible in 2021, allowing closure phase measurements and the start of non-imaging science. When further funding becomes available to allow the completion of a seven-telescope array, the MROI should usher in a step change in the high-angular-resolution imaging capabilities available to astronomers.

## Into space with SuperBIT

It is possible to circumvent scattering by the Earth's atmosphere and achieve high-resolution imaging by either modelling the atmosphere and correcting for it, or simply by raising a telescope above the atmosphere. For most projects, it is prohibitively expensive to launch a large telescope by rocket. But, as **Richard Massey** (Durham University) pointed out, a giant helium balloon can lift 2 tonne payloads above 99% of the Earth's atmosphere for less than 1% of the cost. Recent advances in superpressure materials technology can make a balloon that stretches to the size of a small football stadium stay aloft for months at a time.

The increase in flight duration resulting from this new technology – from a few days to a few months – has suddenly rekindled worldwide interest in scientific ballooning. NASA, Canadian and French space agencies each run a balloon programme, both as a cheap testbed for space technology and as an end in itself. NASA is pushing its balloon programme as a way to preserve UV and multicolour optical astronomy when the Hubble Space Telescope fails. The UK's world-leading balloon programme was shelved in the 1990s. Balloon-borne research in the UK now falls in the gap between funding agencies: too low for the UK Space Agency and too high for the Science and Technology Facilities Council. Fortunately, the costs are such that it was possible to buy into the new long-duration technology and build first-generation balloon telescopes on a university budget.

SuperBIT (the Superpressure Balloon-borne Imaging Telescope) is a diffraction-limited near-UV and optical telescope built by Durham, Toronto and Princeton universities in collaboration with NASA. During three test-flights between 2015 and 2018, SuperBIT achieved high-resolution imaging with a modern camera whose field of view is 36 times larger than that of the Hubble Space Telescope. After each flight, the telescope has descended by parachute, been upgraded (without

> "Recent advances in superpressure materials can make a balloon stay aloft for months"

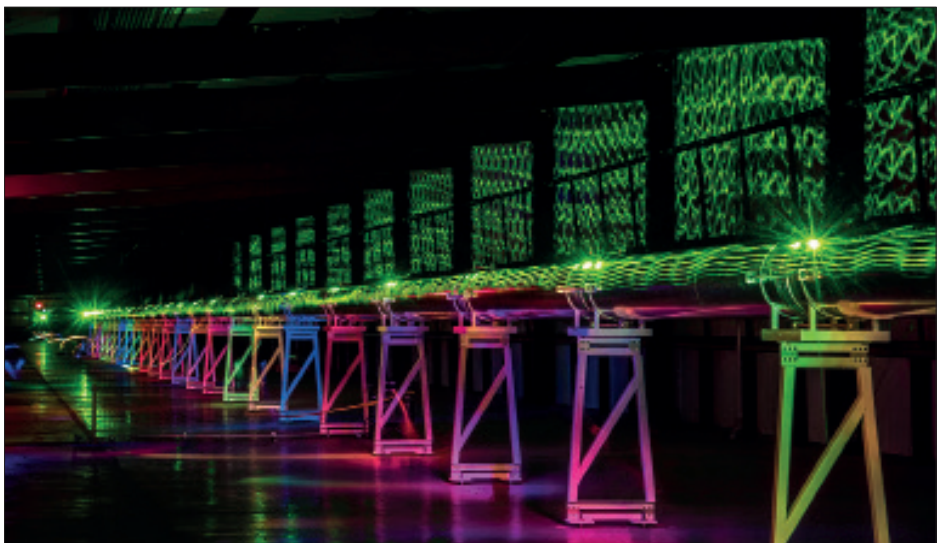

**7** Inside the path compensator building of the Magdalena Ridge Observatory Interferometer (MROI).

the need for a space shuttle), then quickly made ready for relaunch. Its current configuration, with a 0.5 m telescope, is scheduled to be launched on NASA's next long-duration balloon from New Zealand in 2020. The initial science goal is to measure gravitational lensing in around 200 galaxy clusters to map the distribution of dark matter. After that, mirrors up to 2 m are within the balloon's lift capability.

Hanging a telescope under a balloon does bring a few limitations. First, the view at zenith is obscured by the balloon. But elevations down to 20° above the horizon can be observed with negligible airmass, because at 40 km altitude the telescope is effectively in space, with excellent atmospheric transmission (for wavelengths above 300 nm). This means that more of the sky is accessible than from telescopes on the ground; during a typical flight from New Zealand, the telescope could observe anywhere in the Southern Hemisphere or up to 20° north. Secondly, the telescope swings under the balloon, like a giant pendulum. The pointing is stabilized using an upgraded version of hardware from the Spider microwave telescope. Gyroscopes provide coarse stability, to ~1′ rms, then a simple tip/tilt mirror locked to a guide star improves this to ~0.02″ rms. For more information, see Romualdez *et al.* (2018).

Astronomy needs high-resolution optical/IR imaging for many applications, such as searching for life on an exoplanet – an extremely difficult task because of the brightness of the central star around which the planet orbits. Producing high-resolution images and delivering high contrast is much better done from space, but space telescopes are extremely expensive and so much less common than ground-based telescopes. **Ian Parry** (University of Cambridge) described an approach using new technologies – SUPERSHARP Space Telescopes – to develop fast and affordable space telescopes that unfold once they reach orbit. These have the potential to produce much higher resolution than astronomers are ever likely to be able to achieve using conventional telescope structures.

There are several technical challenges in providing such a telescope. First, the unfolding scheme must locate the optical elements with an accuracy in the 2–30 mm range. The system also needs to have a very accurate metrology system to allow the primary mirror segments to be aligned to within about 50 nm. That in turn requires high-accuracy actuators integrated with the precision metrology system and a sophisticated control system. The Cambridge team at the Institute of Astronomy is now assembling a laboratory-based proof-of-concept demonstrator for the unfolding and self-aligning technologies. Then the plan is to find funding for a CubeSat in-orbit demonstrator in the hope that this will lead to scientific and commercial projects.

High-resolution imagery is also important for planetary astrophysics. **Nahid Chowdhury** (University of Leicester) described observations of Saturn's infrared auroral emissions. He used AO observations made with the VLT's long-slit spectrometer, the cryogenic infrared echelle spectrograph (VLT-CRIRES) facility at Paranal, Chile, to re-analyse spectral data from Saturn's northern aurorae. The data revealed a dawn-enhanced auroral intensity profile with a brightening in the aurora that tends towards the dawn sector of the pole – a feature that has been observed in previous ultraviolet and infrared studies and is linked to solar wind compressions in the magnetosphere. He also found evidence for an ionospheric polar vortex that possesses strong noon–midnight ion flows inside the polar cap, which is further linked to a corresponding dark region in the auroral emission intensity at the same location (Geballe *et al.* 1993, Johnson *et al.* 2017, 2018).

The thermospheric temperature profile revealed a subtle and previously undetected increase across the polar cap from the dawn side of the planet to the dusk side. There is an intriguing temperature hotspot on the dusk side of the planet that corresponds to intense emission inside the polar cap; this is evidence for temperature-driven auroral brightening and could help to explain thermospheric flows driving planetary period oscillations (O'Donoghue *et al.* 2014, Stallard *et al.* 2012).

### Conclusions

The meeting was very successful and well attended. What became clear was that there was a great deal of activity in the UK and Europe trying to push the limits of the angular resolution that can be achieved with an exciting variety of novel technologies. High-resolution imaging is already being delivered by a number of instruments on large and medium-sized ground-based telescopes. In most presentations significant new technologies were described at various stages in their development. It became very clear that many of the limits that people imagine are set for us with ground-based telescopes can be overcome with a more creative approach to dealing with our atmosphere and the distortions it creates. We are undoubtedly a long way from the limits of what can be achieved in our imagination and, with appropriate funding, what might be achieved in practice. ●

**AUTHORS**
**David Buscher**, Kavli Institute for Cosmology, Cavendish Laboratory, Cambridge, UK; **Nahid Chowdhury**, Dept of Physics & Astronomy, University of Leicester, UK; **Rick Davies**, Max Planck Institut für extraterrestrische Physik, Garching bei München, Germany; **Sasha Hinkley**, Dept of Physics, University of Exeter, UK; **Norbert Hubin**, European Southern Observatory, Garching bei München, Germany; **Paul Jorden**, Space Astronomy Defense Imaging, Chelmsford, UK; **Craig Mackay**, Institute of Astronomy, University of Cambridge, UK; **Richard Massey**, Institute for Computational Cosmology, Durham University, UK; **Kieran O'Brien**, Centre For Advanced Instrumentation, Durham University, UK; **Ian Parry**, Institute of Astronomy, University of Cambridge, UK; **Jesper Skottfelt**, Centre for Electronic Imaging, Open University, UK.

**REFERENCES**
**Abuter R** *et al.* 2018 *Astron. & Astrophys.* **615** L15
**Buscher D F** *et al.* 2013 *J. Astron. Instrum.* **02(02)** 1340001
**Davies R & Kasper M** 2012 *Ann. Rev. Astron. Astrophys.* **50** 305
**Geballe T** *et al.* 1993 *Astrophys. J.* **408** L109
**Johnson R E** *et al.* 2017 *J. Geophys. Res. Space Phys.* **122** 7599
**Johnson R E** *et al.* 2018 *J. Geophys. Res. Space Phys.* **123** 5990
**Jorden P** *et al.* 2018 *Proc. SPIE* **10709** 2
**Mackay C** *et al.* 2016 *Proc. Astron. Soc. Pac.* **35** 47
**Mackay C** *et al.* 2016 *Proc. SPIE* **99083L**
**O'Donoghue J** *et al.* 2014 *Icarus* **229** 214
**Parry I** *et al.* 2018 SUPERSHARP white paper https://arxiv.org/abs/1801.06111
**Romualdez L J** *et al.* 2018 *Proc. SPIE* **10702**
**Stallard T S** *et al.* 2012 *J. Geophys. Res. Space Phys.* **117** 1